\documentstyle[epsfig]{aipproc}

\begin{document}
\title{First-principles study of \\ lattice instabilities
in Ba$_{x}$Sr$_{1-x}$TiO$_{3}$}

\author{Ph. Ghosez$^*$, D. Desquesnes$^{\dagger}$, X. Gonze$^{\dagger}$ and K. M. Rabe$^{\ddagger}$}
\address{$^*$Institut de Physique, Universit\'e de
Li\`ege, B\^at. B5, B-4000 Li\`ege, Belgium\\
$^{\dagger}$Laboratoire de Physico-Chimie et de Physique des
Mat\'eriaux, \\ Universit\'e Catholique de Louvain, B-1348
Louvain-la-neuve, Belgium \\
$^{\ddagger}$Department of Physics and Astronomy, Rutgers University,
Piscataway, NJ, USA}

\maketitle

\begin{abstract}
Using first-principles calculations based on a variational density
functional perturbation theory, we investigate the lattice dynamics
of solid solutions of barium and strontium titanates. Averaging the
information available for the related pure compounds yields results
equivalent to those obtained within the virtual crystal approximation,
providing frequencies which are a good approximation to those computed for
a (111) ordered supercell. Using the same averaging technique we
report the evolution of the ferroelectric and antiferrodistortive
instabilities with composition.

\end{abstract}

\section*{Introduction}

Ferroelectric solid solutions with perovskite structure can
exhibit properties which are much more interesting for
technological applications than those of the related pure
compounds. This is the case for the new class of single crystal
relaxor ferroelectrics, exhibiting exceptionally high
piezoelectric constants~\cite{Park97}. It is also true for
Ba$_{x}$Sr$_{1-x}$TiO$_{3}$ (BST), which combines a high dielectric
constant with other properties making it a good candidate for
the construction of dynamic random access memories
(DRAM)~\cite{Bilodeau97}.

In contrast to barium titanate (BaTiO$_{3}$) which is a prototype
ferroelectric (FE), strontium titanate (SrTiO$_{3}$) is an incipient
ferroelectric which undergoes an antiferrodistortive (AFD) phase
transition at 105K~\cite{Lines77}. Due to the different
behavior of the related pure compounds, the structural properties of BST
evolve with composition~\cite{Lemanov96}. It exhibits a sequence of 
ferroelectric
transitions similar to that of BaTiO$_{3}$ for Ba concentrations ranging
from 20 to 100\%. Close to pure SrTiO$_{3}$ the behavior is
complicated by the competition between FE and AFD instabilities
yielding the phase diagram reported in Ref. \cite{Lemanov96}.

In the past few years, first-principles effective Hamiltonians
sucessfully described the temperature-dependent behavior of various pure
perovskite oxides~\cite{Vanderbilt97}. The generalization of this
theoretical approach to solid solutions is a new challenge which
includes some additional complications to treat the
alloy disorder accurately.

A straightforward procedure is to try to extract the relevant
information on the alloy from computation for various ordered supercells.
This idea was applied to the study of
Pb$_{x}$Ge$_{1-x}$Te solid solutions~\cite{Cockayne98} but has
the inconvenience of being computationally intensive. An alternative and
promising approach is based on the Virtual Crystal Approximation
(VCA) which allows the retention of a crystal with the primitive periodicity
but composed of virtual atomic potentials averaging those of the
atoms in the parent compounds. A careful construction of the VCA for
PbZr$_{x}$Ti$_{(1-x)}$O$_{3}$ yielded a stress
induced phase transition~\cite{Ramer00} and a compositional phase
boundary~\cite{Ramer01}. Going further, the construction of a model
Hamiltonian based
on VCA calculations for the same compound recently was used to describe
most of its temperature-dependent properties~\cite{Bellaiche00}. In the latter
case, the VCA calculation provided a reference structure to which
corrections had to be added to treat the behavior of the
real ions correctly.

If the purpose of the VCA is to define an average crystal from which
the real solid solution can be described by including a restricted
number of small corrections, one can ask if a similarly good
estimate of the mixed compound could be obtained directly from the
information available for the pure materials by averaging some {\it key}
quantities. This idea is reinforced by the observation that the
interatomic force constants seem very similar in the different
ABO$_{3}$ compounds~\cite{Ghosez99}.

In what follows we define an Average Crystal Approximation (ACA) by
averaging the lattice constant ($a_{cell}$), the Born effective charges
(Z$^{*}$), the optical dielectric tensor ($\epsilon_{\infty}$),
and the interatomic
force constants in real space (IFC) of the pure compounds. We compare
our results to those obtained (i) within the VCA and (ii) for a cubic
ordered supercell with alternating planes of Ba and Sr along the (111)
direction. We show that our ACA yields an excellent approximation to the
dielectric and dynamical properties of the mixed compound. It allows
the prediction of the lattice dynamics in the full range of
composition without requiring any first-principles calculations
other than those for the related pure compounds.

\section*{Methodology}

\subsection*{General framework}

Our calculations have been performed within the density functional
theory~\cite{Jones89} and the local density approximation, using the
{\sc abinit} package~\cite{abinit}. The ionic potentials
screened by the core electrons have been replaced by highly transferable
extended norm conserving pseudopotentials as proposed by M.
Teter~\cite{Teter93}.
The Ba 5s, Ba 5p, Sr 4s, Sr 4p, Ti 3s, Ti 3p, Ti 3d, Ti 4s, O 2s and
O 2p electrons have been treated as valence states. The wavefunction has
been expanded in plane waves up to a kinetic energy cutoff of 45 Ha.
The electronic exchange-correlation energy  has been computed using a
polynomial parametrization\cite{Goedeker96} of Ceperley Alder
data~\cite{Ceperley80}.
Integrals over the simple cubic Brillouin zone have been
replaced by sums on a $6 \times 6 \times 6$ mesh of special $k$-points.
A grid providing an equivalent sampling has been used for the supercell
calculations.

The Born effective charges, the dielectric constants and the dynamical
matrices have been computed within a variational formulation of the
density functional perturbation theory~\cite{Gonze92}. The interatomic force
constants in real space (IFC) and the phonon dispersion curves have been
extracted following the scheme described in Ref. \cite{Gonze94}, which deals
correctly with the long range character of the dipolar interaction.
The IFC were obtained from the knowledge of the dynamical matrix on a
$2 \times 2 \times 2$ BCC mesh of $q$-points.

\subsection*{Virtual atom pseudopotential}

The virtual crystal approximation is based on the construction of a
fictitious ``virtual atom'' potential by averaging the ionic potentials
of the atoms alternating at the same site of the structure. As
illustrated recently by Ramer and Rappe~\cite{Ramer00}, there are 
different ways of
combining non-local atomic pseudopotentials, each of them leading to
different physical results. A careful choice seems necessary to predict
accurately the alloy properties at the VCA level. However, this choice could
be less stringent if we think about the virtual crystal as a 
reference structure
from which the real crystal can be obtained by including a limited number
of relevant corrections.

For BST, we choose therefore to stay with the most straightforward
combination of the reference atomic pseudopotentials.
The potential at the A-site can be formally defined for a
concentration of $x$\% of barium as:
\begin{equation}
     \label{eq-vca}
     V_{VCA}^{ps}[x]= x V_{Ba}^{ps} + (1-x) V_{Sr}^{ps}
\end{equation}
where $V_{Ba}^{ps}$ and $V_{Sr}^{ps}$ are the pseudopotentials of Ba
and Sr atoms.

Non-local pseudopotentials, such as those we are using,
  can be expressed as a sum of a local
contribution and short-range non-local corrections:
\begin{equation}
V^{ps}(r,r') = V^{loc}(r) \delta(r-r')
+ \sum_{l,m} \sum_{q}
\frac{ V^{ps}_{l} | \phi_{lm,q}> <\phi_{lm,q} | V^{ps}_{l}}
      {< \phi_{lm,q} | V^{ps}_{l} | \phi_{lm,q} > }.
\end{equation}
Our virtual atom pseudopotential is then defined as
follows~\cite{Desquesnes}:
\begin{eqnarray}
V^{ps}_{VCA}(r,r') &=& [x V^{loc}_{Ba}(r) + (1-x) V^{loc}_{Sr}(r)]
\delta(r-r') \nonumber\\
&+& \sum_{l,m} \sum_{q}
\frac{ x V^{ps}_{Ba,l} | \phi^{Ba}_{lm,q}> <\phi^{Ba}_{lm,q} | x V^{ps}_{Ba,l}}
      {< \phi^{Ba}_{lm,q} | x V^{ps}_{Ba,l} | \phi^{Ba}_{lm,q} > }
      \nonumber \\
&+&  \sum_{l,m} \sum_{q}
\frac{ (1-x) V^{ps}_{Sr,l} | \phi^{Sr}_{lm,q}> <\phi^{Sr}_{lm,q} |
(1-x) V^{ps}_{Sr,l}}
      {< \phi^{Sr}_{lm,q} | (1-x) V^{ps}_{Sr,l} | \phi^{Sr}_{lm,q} > }
\end{eqnarray}
For $x$ ranging from 0 to 1, our virtual atom evolves smoothly from
Sr to Ba.

\section*{Ground-state properties}

First we focus on the lattice constant of the cubic perovskite
structure and look at its evolution with composition.
In Table~\ref{Table-GS}, we report the LDA lattice constants 
for BaTiO$_3$ and SrTiO$_3$, for Ba$_{1-x}$Sr$_x$TiO$_3$ within 
the VCA (for x = 0.25, 0.50 and 0.75), and for the (111) ordered 
supercell. We compare the computed values with experimental results, 
which at intermediate compositions are accurately given by Vegard's 
law~\cite{Lemanov96}. 
While our results show the lattice constant underestimate
typical of LDA, the Vegard's law behavior is well reproduced.

\begin{table}[tbm]
\caption{Evolution of the LDA optimized lattice constant (bohr) with
composition. Experimental values on the third line are interpolated
from the data in the pure compounds using Vegard's law.}
\label{Table-GS}
\begin{tabular}{lcccccc}
  &SrTiO$_{3}$ & x= 25\% & x= 50\% & x= 75\% &BaTiO$_{3}$
     \\
\tableline
LDA/VCA   &7.273   &7.321   &7.370  &7.412  &7.451     \\
LDA/Supercell&     &        &7.366  &       &           \\
Experiment  &7.38    &7.43    &7.47   &7.52   &7.56       \\
\end{tabular}
\end{table}

We note that the theoretical LDA underestimate of the
experimental lattice constant fortuituously  corresponds to the 
effect of a similar external pressure of 8.7 GPa in both 
BaTiO$_{3}$ and SrTiO$_{3}$.
In what follows, we will work at the experimental lattice constants
which is therefore equivalent to imposing a fixed fictitious negative
pressure of -8.7 GPa to the LDA results. For the mixed compounds we
will work at the values obtained by Vegard's law as reported on the
third line of Table~\ref{Table-GS}.

\section*{Dielectric and dynamical properties}

\subsection*{Born effective charges and dielectric tensor}

In Table~\ref{Table-Z}, we compare the Born effective charges and the
macroscopic optical dielectric constants obtained in our various
calculations. The values of O$_{\parallel}$ and O$_{\perp}$ refer
respectively to the change of polarization for a displacement of the
oxygen atom along the Ti--O direction and perpendicular to it.

\begin{table}[tb]
\caption{Born effective charges and optical dielectric constant in the
cubic perovskite structure. In the supercell, we can have either Ba
or Sr at the A site: the first value refers to Ba, the second to Sr.}
\label{Table-Z}
\begin{tabular}{lccccc}
Atom & BaTiO$_{3}$ & SrTiO$_{3}$ &Average &VCA &Supercell  \\
\tableline
Z$^{*}_{A}$              &$+2.74$  &$+2.55$  &$+2.64$  &$+2.63$ 
&$+2.71,+2.58$ \\
Z$^{*}_{B}$              &$+7.32$  &$+7.26$  &$+7.29$  &$+7.29$  &$+7.29$ \\
Z$^{*}_{O_{\parallel}}$  &$-5.78$  &$-5.74$  &$-5.76$  &$-5.75$  &$-5.76$\\
Z$^{*}_{O_{\perp}}$      &$-2.14$  &$-2.04$  &$-2.09$  &$-2.08$  &$-2.09$\\
$\epsilon_{\infty}$  &$6.75$   &$6.33$   &$6.54$   &$6.51$   &$6.54$ \\
\end{tabular}
\end{table}

We observe that the values of Z$^{*}$ and $\epsilon_{\infty}$ 
in the supercell are very
similar to both those averaged from the pure compounds (ACA) and
those obtained in the VCA. In particular, the ACA and the VCA similarly
reproduce the large anomalous values of $Z^{*}_{Ti}$ and $Z^{*}_{O \parallel}$
which play an important role in generating the FE instability.
For Z$^{*}_{A}$, both kinds of approximation yield the average of
the values computed for Ba and Sr in the supercell. A similar observation
was reported by Bellaiche {\it et al.} for PZT~\cite{Bellaiche00}.

\subsection*{Phonon frequencies and interatomic force constants}

Finally we compare the phonon dispersion curves, yielding information
at the harmonic level on the full lattice dynamics of the crystal.
Phonon frequencies are obtained by diagonalizing the dynamical matrix.
The latter is usually decomposed into a dipolar part computed from
the values of Z$^{*}$ and $\epsilon_{\infty}$, and a short-range part
computed using the IFC. Our ACA dynamical matrix is constructed by simply averaging
all the previous quantities from their values in the pure compounds
and using an average mass at the A-site, while the VCA requires a full
set of additional first-principles
calculations with the virtual atom potential.

\begin{figure}
~\mbox{~}
\hspace{2cm}
\epsfig{file=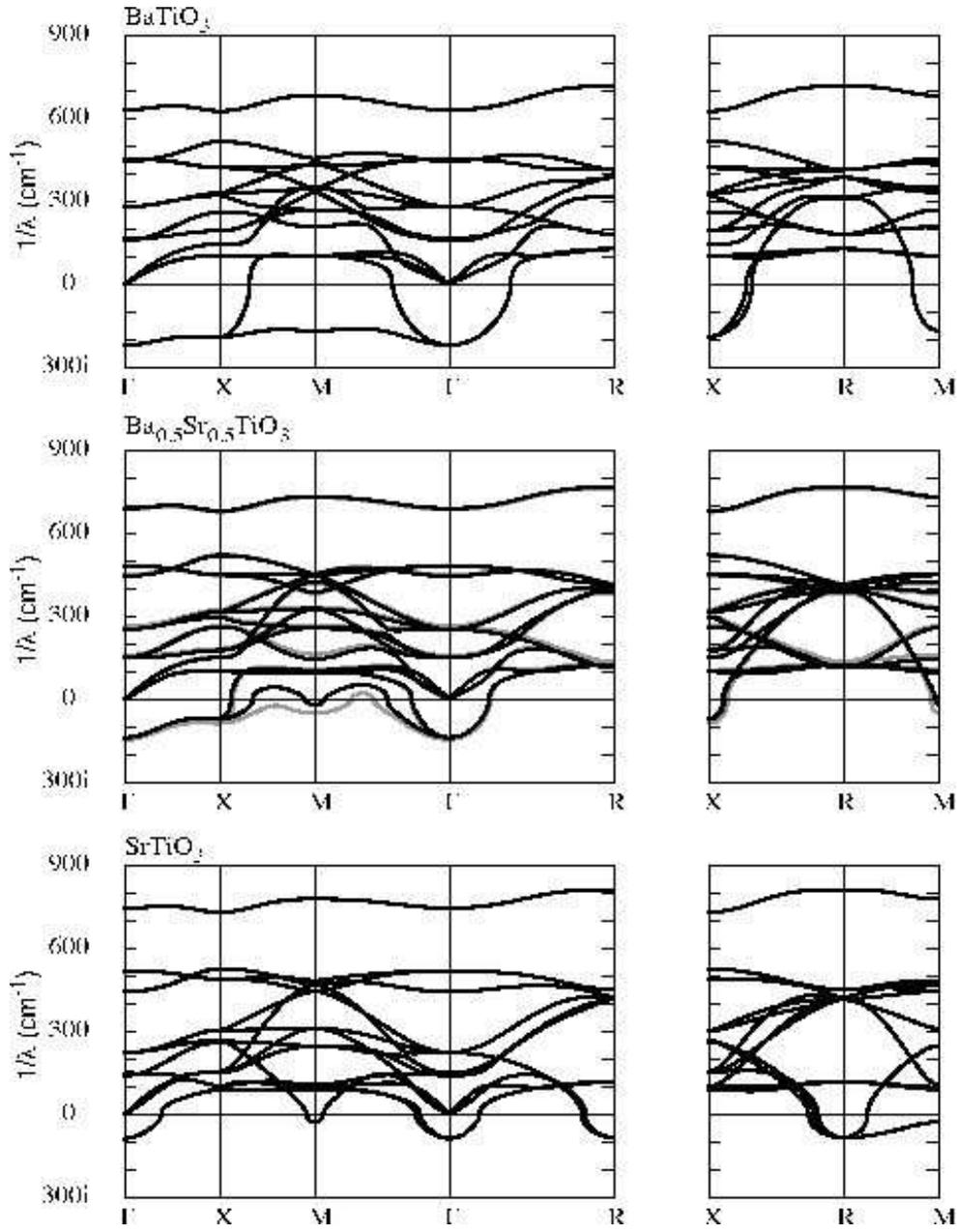,width=13cm}
\vspace{1cm}
\caption{Phonon dispersion curves along high symmetry lines of the
simple cubic Brillouin zone for BaTiO$_{3}$,
Ba$_{0.5}$Sr$_{0.5}$TiO$_{3}$ and SrTiO$_{3}$. In the middle panel,
the black lines correspond to the ACA, and the grey line to the VCA.}
\label{Fig-1}
\end{figure}

In Fig.~\ref{Fig-1}, we show the phonon dispersion curves of a mixed
crystal of BST with x=50\%, obtained both within the ACA and the VCA.
We compare the results to those obtained for the related pure compounds.
The dispersion curves of BaTiO$_{3}$ were discussed in Ref. \cite{Ghosez98}.
Those of SrTiO$_{3}$ have been recomputed and agree with the
calculation reported previously by LaSota {\it et al.} \cite{LaSota97}.
Compared to BaTiO$_{3}$ which exhibits only a FE instability with a flat
dispersion in the $\Gamma$--X--M plane, the FE instability in SrTiO$_{3}$
is more localized in $q$-space (around the $\Gamma$ point) and coexists
with an AFD instability extending along the Brillouin zone edges
(R--M lines).

We observe that averaging the different quantities (Z$^{*}$,
$\epsilon_{\infty}$, IFC, atomic masses) does not yield an average of the
phonon frequencies but rather is  
closer to the results for pure
BaTiO$_{3}$. This is also what is found in the VCA, the two phonon dispersions
being very similar.
The major discepancy appears for the low frequency Ti
dominated branch along the $\Gamma$--M--X lines
where the value of $\omega^2$, being near zero, is especially sensitive to
small
changes in the dynamical matrix.

\begin{table}[tb]
\caption{Phonon frequencies (cm$^{-1}$) at the $\Gamma$ and R points of
the simple cubic Brillouin zone. Comparison of the ACA and VCA results
with the supercell eigenmode at the $\Gamma$ point which has the highest
overlap with the corresponding ACA eigenvector.}
\label{Table-w}
\begin{tabular}{lccccc}
Mode & BaTiO$_{3}$ & SrTiO$_{3}$ &Average &VCA &Supercell \\
\tableline
$\Gamma_{15}$(TO)    &$219i$ &$87i$  &$138i$ &$145i$ &$151i$ \\
$\Gamma_{15}$(A)     &$0$    &$0$    &$0$    &$0$    &$0$   \\
$\Gamma_{15}$(LO)    &$159$  &$141$  &$152$  &$158$  &$160$ \\
$\Gamma_{15}$(TO)    &$166$  &$149$  &$152$  &$159$  &$161$ \\
$\Gamma_{25}$        &$281$  &$223$  &$253$  &$263$  &$272$ \\
$\Gamma_{15}$(LO)    &$445$  &$443$  &$446$  &$449$  &$450$ \\
$\Gamma_{15}$(TO)    &$453$  &$519$  &$482$  &$481$  &$480$ \\
$\Gamma_{15}$(LO)    &$631$  &$746$  &$690$  &$687$  &$687$ \\
\hline
$R_{25}$             &$182$  &$84i$  &$135$  &$114$  &$86$\\
$R_{15}$             &$128$  &$118$  &$131$  &$125$  &$110-111$ \\
$R_{12'}$            &$314$  &$451$  &$384$  &$392$  &$384$ \\
$R_{25'}$            &$386$  &$418$  &$405$  &$406$  &$404$ \\
$R_{15}$             &$414$  &$423$  &$419$  &$415$  &$412-418$\\
$R_{2'}$             &$717$  &$812$  &$764$  &$766$  &$763$ \\
\end{tabular}
\end{table}

In Table~\ref{Table-w}, we compare the ACA and VCA phonon frequencies
at $\Gamma$ and R in the reference perovskite structure to the related
frequencies at the $\Gamma$ point in the (111) ordered
supercell. The different symmetry of the supercell does not allow a
one to one correspondence but the comparison is made between phonons
having the highest eigenvector overlap. We note that the supercell
eigenmode associated with the R$_{15}$ modes of the perovskite structure,
and involving displacement of the A atom, exhibits a small LO-TO
splitting.

The overall agreement between ACA, VCA and supercell calculations is
very good. This is particularly true for the $\Gamma_{15}$ and
R$_{25'}$ modes related to the FE instability. There is a larger
discrepancy for the R$_{25}$ AFD mode. We note however that projecting
the supercell dynamical matrix within the subspace of the VCA
eigenvector yields a value of 125 cm$^{-1}$ (instead of 86
cm$^{-1}$) closer to VCA and ACA results. This shows that the
discrepancy originates in a change of the eigenvector rather than in a
modification of the interatomic force constants.

\subsection*{Evolution of the phonon instabilities with composition}

\begin{figure}[tb]
~\mbox{~}
\hspace{2cm}
\epsfig{file=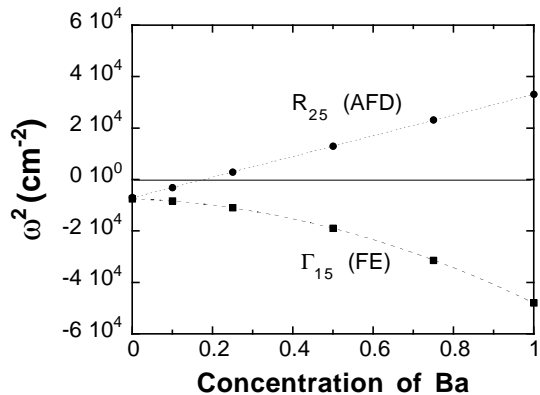,width=7cm}
\vspace{1cm}
\caption{Evolution of the R$_{25}$ and $\Gamma_{15}$ unstable mode
with composition, using the ACA.}
\label{Fig-2}
\end{figure}

As the ACA seems to give accurate results for the lattice dynamics of
BST solid solutions, it offers an easy way to make a guess for the
evolution of the FE and AFD instabilities with composition. We observe in
Figure \ref{Fig-2} that the square of the frequency of the R$_{25}$
AFD mode evolves linearly with composition and becomes unstable for
Ba concentrations smaller than 20 $\%$. This corresponds precisely
to the composition for which a change of behavior is observed experimentally
in the phase diagram of BST~\cite{Lemanov96}. In contrast, the FE instability
exhibits a non-linear behavior and remains unstable for the full range of
composition.

\section*{Conclusions}
In this paper, we have compared different approaches for the study of
the lattice dynamics of BST solid solutions and have introduced a new
Average Crystal Approximation. We have shown that the latter yields
results equivalent to the VCA but with the advantage that it does not
require any additional first-principles calculations if
the information for the related pure compounds is available. Looking
at the actual values of the IFC in the supercell could
provide indications of the corrections to be included in the harmonic
part of an effective Hamiltonian based on the ACA for the correct description of
BST solid solutions.

\section*{Acknowledgments}
This work was supported by ONR grant No N00014-97-0047. Supercomputing 
support was provided by MHPCC. X.G. acknowledges financial support from 
the F.N.R.S. (Belgium), the F.R.F.C. (project 2.4556.99), the P.A.I. P4/10.

\end{document}